\documentclass[11pt]{article}

\usepackage[a4paper,margin=25mm]{geometry}
\usepackage{graphicx}
\usepackage{amsmath,amssymb}
\usepackage[T1]{fontenc}
\usepackage{lmodern}
\usepackage{authblk}
\usepackage{url}

\usepackage{tikz}
\usepackage[authoryear,round,sort&compress]{natbib}

\usepackage[hidelinks,colorlinks=true,citecolor=blue,urlcolor=blue,linkcolor=blue]{hyperref}

\DeclareRobustCommand{\ORCIDicon}[1]{%
  \,\texorpdfstring{%
  \href{https://orcid.org/#1}{%
    \begin{tikzpicture}[baseline=-0.6ex]
      \node[inner sep=0pt] {\includegraphics[height=1.6ex]{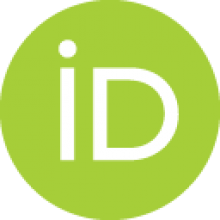}};
    \end{tikzpicture}%
  }}{}%
}

\title{PyIRD: A Python-Based Data Reduction Pipeline for Subaru/IRD and REACH}

\author[1]{Yui Kasagi\ORCIDicon{0000-0002-8607-358X}}
\author[1,2]{Hajime Kawahara\ORCIDicon{0000-0003-3309-9134}}
\author[2]{Ziying Gu}
\author[3,4,5]{Teruyuki Hirano\ORCIDicon{0000-0003-3618-7535}}
\author[3,4,5]{Takayuki Kotani\ORCIDicon{0000-0001-6181-3142}}
\author[3,5]{Masayuki Kuzuhara\ORCIDicon{0000-0002-4677-9182}}
\author[6]{Kento Masuda\ORCIDicon{0000-0003-1298-9699}}

\affil[1]{Institute of Space and Astronautical Science, JAXA, Sagamihara, Kanagawa 252-5210, Japan}
\affil[2]{Department of Astronomy, Graduate School of Science, The University of Tokyo, Tokyo 113-0033, Japan}
\affil[3]{Astrobiology Center, 2-21-1 Osawa, Mitaka, Tokyo 181-8588, Japan}
\affil[4]{Astronomical Science Program, SOKENDAI, 2-21-1 Osawa, Mitaka, Tokyo 181-8588, Japan}
\affil[5]{National Astronomical Observatory of Japan, 2-21-1 Osawa, Mitaka, Tokyo 181-8588, Japan}
\affil[6]{Department of Earth and Space Science, Osaka University, Toyonaka, Osaka 560-0043, Japan}

\date{17 January 2026}

\begin{document}

\maketitle

\section*{Summary}
\label{summary}

\texttt{PyIRD} is a Python-based pipeline for reducing spectroscopic
data obtained with IRD (InfraRed Doppler; \citet{kotani2018}) and REACH
(Rigorous Exoplanetary Atmosphere Characterization with High dispersion
coronagraphy; \citet{kotani2020}) on the Subaru Telescope. It is designed to
process raw images into one-dimensional spectra in a semi-automatic
manner. Unlike traditional methods, it does not rely on \texttt{IRAF}
\citep{tody1986, tody1993}, a software used for astronomical data
reduction. This approach simplifies the workflow while maintaining
efficiency and accuracy. Additionally, the pipeline includes an updated
method for removing readout noise patterns from raw images, enabling
efficient extraction of spectra even for faint targets such as brown
dwarfs.

The code is open source and available at \url{https://github.com/prvjapan/pyird}.

\section*{Statement of need}
\label{statement-of-need}

The reduction of high-dispersion spectroscopic data has traditionally
been performed using \texttt{IRAF}, one of the most widely used tools
for astronomical data reduction and analysis. Although the National
Optical Astronomy Observatory (NOAO) officially ceased development and
maintenance of \texttt{IRAF} in 2013, community-based maintenance has
continued. However, the official IRAF community distribution\footnote{IRAF
  Community Distribution website: \url{https://iraf-community.github.io/}} and
the Space Telescope Science Institute (STScI)\footnote{STScI Newsletter
  (2018), ``Removing the Institute's Dependence on IRAF: You Can Do It
  Too'':
  \url{https://www.stsci.edu/contents/newsletters/2018-volume-35-issue-03/removing-the-institutes-dependence-on-iraf-you-can-do-it-too}}
have both recommended that researchers transition away from
\texttt{IRAF} due to its legacy architecture and lack of institutional
support.

In recent years, several open-source, Python-based pipelines for the
reduction of near-infrared echelle spectrographs have been developed.
Some pipelines utilize \texttt{PyRAF}, a Python interface to IRAF, such
as \texttt{WARP} for the WINERED spectrograph \citep{Hamano2024}, while
others, including \texttt{PLP} for IGRINS \citep{Sim2014} and
\texttt{PypeIt} \citep{pypeit:joss_pub}, do not rely on PyRAF or
IRAF-based components. While these pipelines provide either general
frameworks or instrument-specific solutions, \texttt{PyIRD} is designed
to offer a simple pipeline optimized for IRD and REACH data reduction.
Furthermore, recent advances combining adaptive optics with these
instruments have enabled high-dispersion spectroscopic observations of
faint companions orbiting bright main-sequence stars. To support such
observations, \texttt{PyIRD} implements improved detector noise
reduction to extract high-quality spectra from these faint targets.

Together, these developments underscore the need for actively
maintained, scalable, and flexible software for high-dispersion
spectroscopic data reduction. \texttt{PyIRD} addresses this need by
providing a modern, Python-based pipeline and has already been utilized
in several studies \citep{kasagi2025, kawashima2025, kawahara2025,tomoyoshi2024}.

\section*{Key Features}
\label{key-features}

\begin{figure}
  \centering
  \includegraphics[keepaspectratio, width=0.9\linewidth]{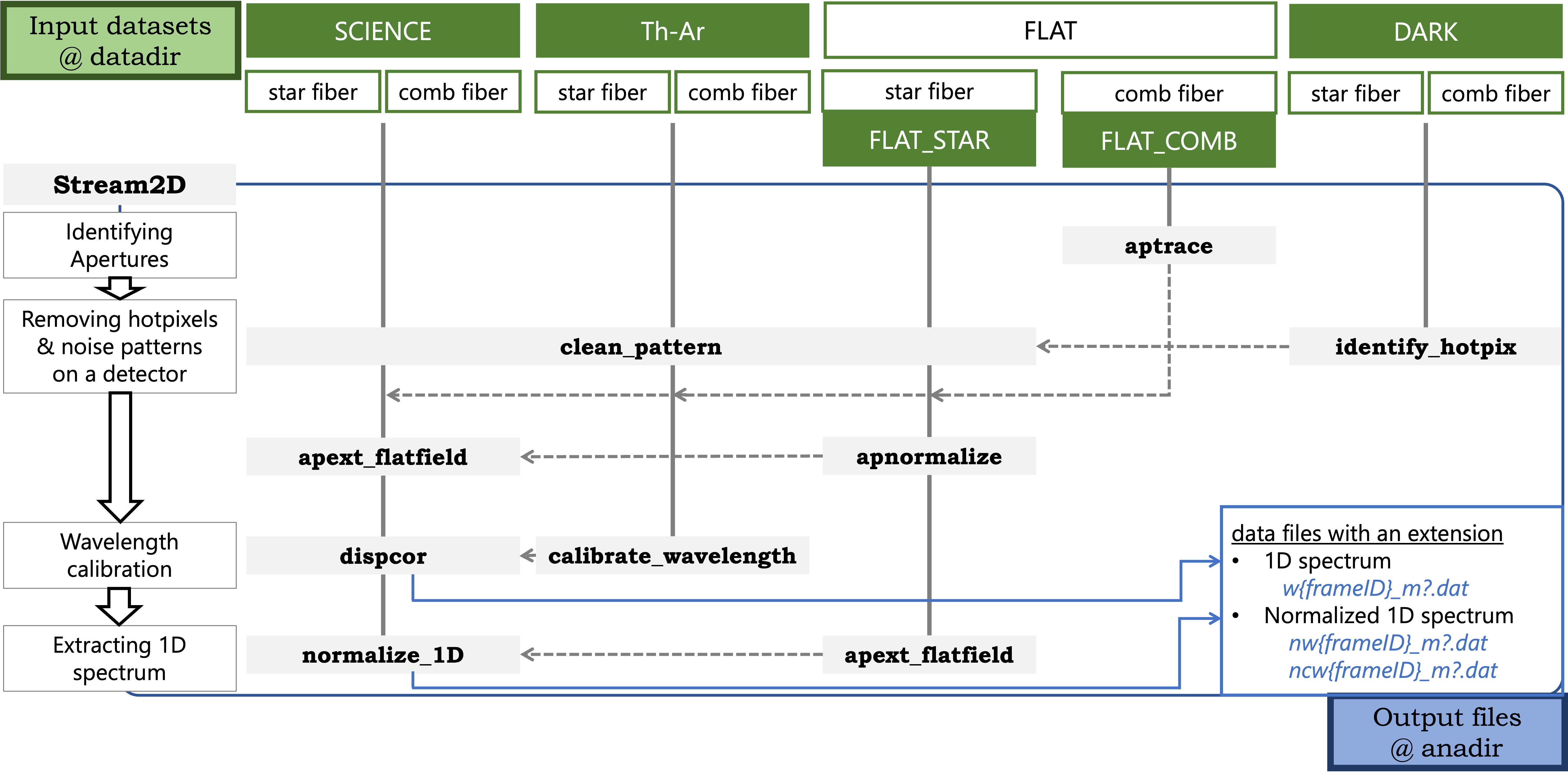}
  \caption{Flowchart of the reduction process for IRD and REACH data. The
  reduction process follows from top to bottom of this figure. Text in the
  grey boxes represents the instance names of each reduction step used in
  \texttt{PyIRD}. \label{fig:reduc_flow}}
\end{figure}

\texttt{PyIRD} performs semi-automatic data reduction by following a
general workflow for high-dispersion spectroscopic data, as illustrated
in \autoref{fig:reduc_flow}. It primarily utilizes \texttt{Astropy}
\citep{astropy:2022}, \texttt{NumPy} \citep{harris2020array},
\texttt{SciPy} \citep{2020SciPy-NMeth}, and \texttt{pandas}
\citep{reback2020pandas}.

To simplify the handling of a large number of input FITS-format files,
\texttt{PyIRD} introduces a Python class called \texttt{FitsSet}. Once
initialized with parameters such as the file IDs and the directory
containing those files, \texttt{FitsSet} automatically organizes and
manages the input files and their metadata. It also allows users to
apply reduction functions collectively to a specified list of FITS IDs,
enabling efficient and consistent data processing through the
\texttt{Stream2D} class.

Since all functions in \texttt{PyIRD} are written in Python rather than
IRAF's Subset Preprocessor Language (SPP), the package is easy to
develop and maintain. This also significantly reduces the time required
for the reduction process: users only need to execute a single Python
script without complex IRAF configuration. For example, reducing data
with \texttt{PyIRD} typically takes several tens of minutes to produce
one-dimensional spectra from raw data obtained during a single observing
night, compared to approximately half a day with traditional IRAF
methods.

\begin{figure}
  \centering
  \includegraphics[keepaspectratio, width=0.9\linewidth]{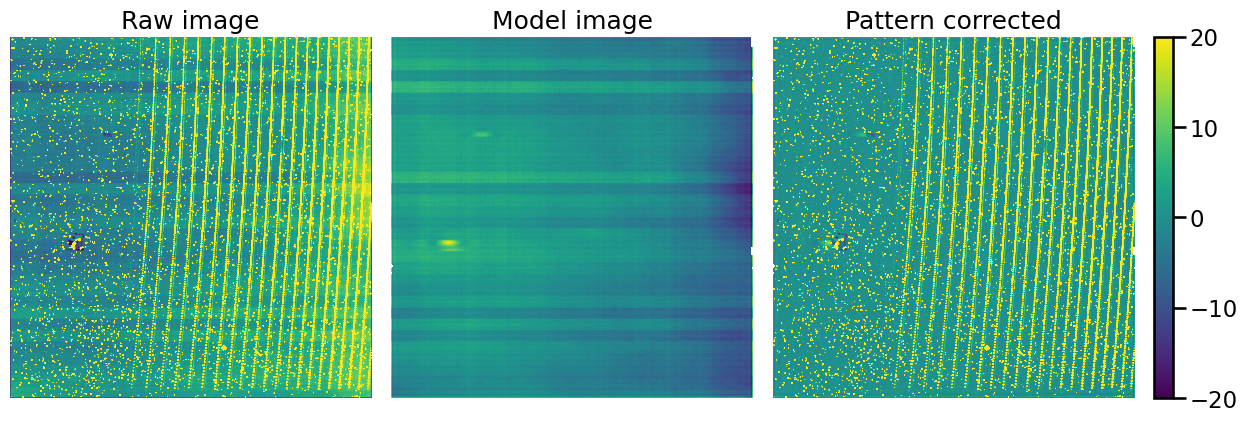}
  \caption{(Left) Raw image; (Middle) Readout pattern model created by
  \texttt{PyIRD}; (Right) Pattern-corrected image \label{fig:pattern}}
\end{figure}

Moreover, \texttt{PyIRD} achieves a higher level of readout noise
pattern removal in the final reduced data. This feature is particularly
important for processing data from faint objects such as brown dwarfs,
where the astronomical signal is often comparable in strength to
systematic noise. The dominant noise source is the readout pattern from
the H2RG detector used in IRD. To address this, \texttt{PyIRD} models
the noise by calculating a median profile for each readout channel and
applying a 2D Gaussian Process using \texttt{gpkron} \citep{gpkron2022}.
This innovative method effectively mitigates the readout pattern, as
shown in \autoref{fig:pattern}, and improves data quality for faint
targets.

\section*{Acknowledgements}
\label{acknowledgements}

Y.K. acknowledges support from JST SPRING, Grant Number JPMJSP2104 and
JSPS KAKENHI grant No.~24K22912. Z.G. acknowledges support from
Forefront Physics and Mathematics Program to Drive Transformation
(FoPM), a World-leading Innovative Graduate Study (WINGS) Program, the
University of Tokyo.

\bibliographystyle{plainnat}
\bibliography{paper}

\end{document}